\documentclass{elsart}

\usepackage{graphicx}
\usepackage{amssymb}

\begin{document}

\begin{frontmatter}

% Title, authors and addresses

% use the thanksref command within \title, \author or \address for footnotes;
% use the corauthref command within \author for corresponding author footnotes;
% use the ead command for the email address,
% and the form \ead[url] for the home page:
% \title{Title\thanksref{label1}}
% \thanks[label1]{}
% \author{Name\corauthref{cor1}\thanksref{label2}}
% \ead{email address}
% \ead[url]{home page}
% \thanks[label2]{}
% \corauth[cor1]{}
% \address{Address\thanksref{label3}}
% \thanks[label3]{}

\title{Electron-hole bilayer quantum dots: Phase diagram and exciton localization}

% use optional labels to link authors explicitly to addresses:
% \author[label1,label2]{}
% \address[label1]{}
% \address[label2]{}

\author{}
\author[inst1]{K. K\"arkk\"ainen} 
\author[inst1]{M. Koskinen} 
\author[inst1]{M. Manninen}
\author[inst2]{S.M. Reimann}

\address[inst1]{Department of Physics, University of Jyv\"askyl\"a,
FIN-40014, Finland}
\address[inst2]{Mathematical Physics, Lund Institute of Technology, P.O. Box 118, S-22100 Lund, Sweden}

\begin{abstract}
We studied a vertical ``quantum dot molecule'',
where one of the dots is occupied with electrons and the other with holes.
We find that different phases occur in the ground state,
depending on the carrier density and the interdot distance.
When the system is dominated by shell structure, 
orbital degeneracies can be removed either by Hund's rule, or by Jahn-Teller 
deformation. Both mechanisms can lead to a maximum of the addition energy 
at mid-shell. At low densities and large interdot distances, bound 
electron-hole pairs are formed. 
\end{abstract}

\begin{keyword}
% keywords here, in the form: keyword \sep keyword
Quantum dot molecule \sep Electron-hole plasma
% PACS codes here, in the form: \PACS code \sep code
\PACS 73.21.La \sep 71.35.Ee \sep 73.21.-b
\end{keyword}
\end{frontmatter}

% main text
%\section{}
%\label{}

Semiconductor quantum dots share many qualities with natural atoms, 
and are thus often referred to as ``artificial atoms''
(for reviews see \cite{chakraborty1999,reimann2002}). In single quantum dots,
shell structure could be observed~\cite{tarucha1996}: large differences in the 
addition energies occurred at particular electron numbers ($N=2, 6, 12$)
corresponding to closed shells, forming the ``quantum dot noble gases''. 
Just like in real atoms, for half-filled shells
spin alignment due to Hund's rule leads 
to an increased stability of the system~\cite{tarucha1996,koskinen1997}.
Quantum dots are uniquely suited to explore regimes of many-electron systems 
that are not accessible in conventional atomic physics, and
much fundamental new insight on correlated states in these 
low-dimensional many-body systems was gained in the past years. 

Quantum dot artificial atoms can be coupled to form ``artificial molecules''.
Such systems have been studied much both experimentally~\cite{doubledotexp}
and theoretically~\cite{doubledotth}, and were even discussed in connection with 
potential quantum computing devices~\cite{quantcomp}.
Most work focused on vertically or laterally coupled double 
quantum dots with electrons. More recently, 
Anisimovas and Peeters~\cite{anisimovas2002a,anisimovas2002b} 
examined the correlated
few-particle states in a vertical {\it bipolar} quantum dot molecule,
consisting of two vertically coupled quantum dots, one populated by 
electrons, the other by holes. They applied the exact diagonalization 
method in the limit of a strong, spin-polarizing magnetic 
field~\cite{anisimovas2002a}
and furthermore studied the dynamic response within a hydrodynamic 
model~\cite{anisimovas2002b}.
The experimental realization of such electron-hole double quantum dots 
should for example be possible in 
bilayer-bipolar heterostructures with separated electron- and hole layers 
in equilibrium, as in biased GaAs/AlGaAs
or InAs/GaSb~\cite{sivan1992}. 
These bipolar systems have attracted much 
attention recently, last but not least due to
the exciting possibility of making a Bose-Einstein condensate with indirect 
excitons~\cite{butov2002}. The phases of a symmetric electron-hole bilayer 
system were recently also investigated by Liu {\it et al.}~\cite{liu1996},
and by Paolo {\it et al.}~\cite{depaolo2002} in the fixed-node 
diffusion Monte Carlo 
approach~\cite{dmc} with periodic boundary conditions. 

In this Communication, we demonstrate that small bipolar-bilayer systems 
consisting of only a few electrons and holes show a very rich phase diagram 
which in fact appears to be  
different from the ``bulk''~\cite{depaolo2002}: depending on the 
electron- and hole-densities and the interdot distance, 
the ground state is either dominated by shell 
structure, leading to a competition between 
Jahn-Teller deformation~\cite{jahn1937} and spin alignment 
in the case of orbital degeneracies, or by the formation of localized, bound 
electron-hole pairs (``localized excitons'').

\begin{figure}
\includegraphics[width=1.0\columnwidth]{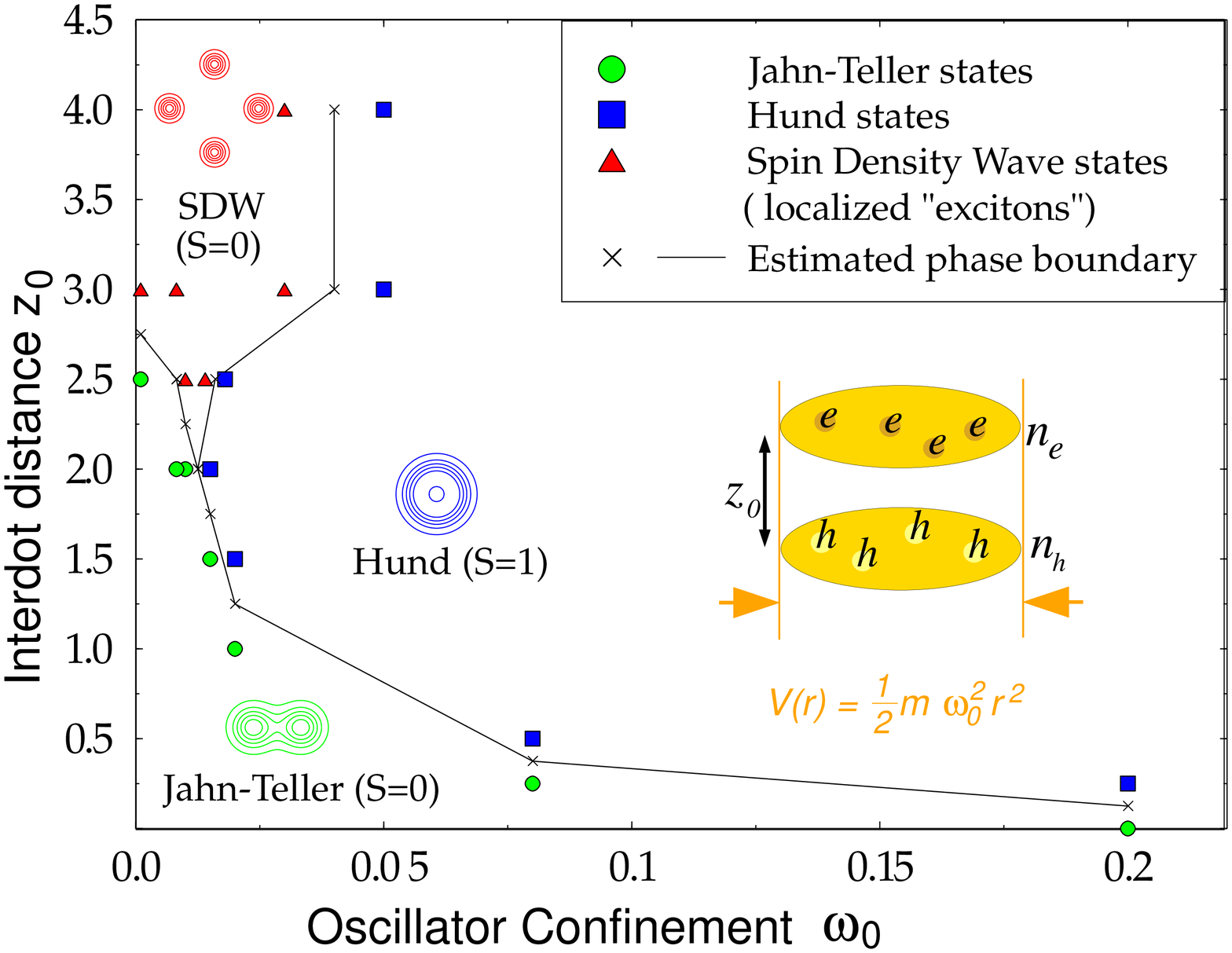}
\caption{Phase diagram of an electron-hole double dot with 4 electrons and 4 holes.
$\omega_0$ is the strength of the external confinement and $z_0$ the interdot distance
(in atomic units).
The dashed lines show the estimated phase boundaries separating different spin-structures.
The characteristic particle density is shown for each phase.
The squares (spin $S=1$) and triangles ($S=0$) indicate the calculated points which
are closest to the estimated phase boundaries.
{\it Inset to the right:} 
Schematic picture of the bipolar quantum dot molecule with 
electrons $(e)$ and holes $(h)$ harmonically confined in two layers, which 
are separated by a distance $z_0$.
}
\label{fig:kimmof1}
\end{figure}

As schematically sketched in the inset to Fig.~\ref{fig:kimmof1},
a quantum dot with electrons in one (two-dimensional) layer is 
separated by a distance 
$z_0$ from a quantum dot confining holes in 
the other 2D layer. The charge carriers 
in both layers are confined by a harmonic potential, 
$V=m_{e,h}^*\omega _0^2r^2/2$, with
$r^2=x^2+y^2$ (a model which is frequently applied to describe single 
quantum dots~\cite{shikin1991,kumar1990}).
The two layers are electrostatically coupled with no interdot tunneling, i.e., 
recombination is prohibited. 

To obtain the ground state energies and densities, we apply 
the density functional method in the local spin density 
approximation~\cite{kohnsham}. For the bipolar bilayer system, we have 
to solve four coupled equations for the electron- and hole spin densities,
$n^e_{\uparrow }({\bf r}), n^e_{\downarrow }({\bf r})$ and   
$n^h_{\uparrow }({\bf r}),  n^h_{\downarrow }({\bf r})$. 
For electrons and holes, respectively, the 
effective Kohn-Sham potential $v_{\rm eff} ^{e(h)\sigma }({\bf r})$ 
consists of the external
harmonic confinement, the electron-electron (hole-hole)
repulsion, the electron-hole (hole-electron) attraction and
the exchange-correlation potential $v_{xc}^\sigma$ which is approximated in the local spin-density 
approximation. We use the von Barth and Hedin ~\cite{barth1972} 
formulation of the local electron (hole) 
exchange-correlation energy $\varepsilon_{xc}(n^{e(h)},\zeta^{e(h)})=\varepsilon_{xc}(n^{e(h)},0)+
f(\zeta^{e(h)})[\varepsilon_{xc}(n^{e(h)},1)-\varepsilon_{xc}(n^{e(h)},0)]$, where $\zeta^{e(h)}$ 
is the electron (hole)
spin polarization and $f(\zeta^{e(h)})=((1+\zeta^{e(h)})^{3/2}+(1-\zeta^{e(h)})^{3/2}-2)/(2^{3/2}-2)$ 
is the polarization 
dependence. The function $f(\zeta^{e(h)})$ 
interpolates between the paramagnetic 
($\zeta^{e(h)}=0$) and ferromagnetic ($\zeta^{e(h)}=1$) cases given by Tanatar and 
Ceperley \cite{tanatar1989}.
The electron-hole correlation is neglected. This is a good approximation for
large interdot separation (the correlation energy is then diminished) 
and for strong external confinement (the correlation energy is then small compared to the
kinetic and exchange energies).
Even in the equilibrium density of the
electron hole plasma the correlation energy is much smaller
than the exchange energy~\cite{brinkman1973} (the equilibrium density 
minimizes the total energy of the two-dimensional plasma in an infinite
system).

For example,
the Kohn-Sham single particle wave functions for the electrons in 
one of the two dots,
$\psi^e_{i\sigma}({\bf r})$, satisfy the Kohn-Sham equations with
the effective potential for the electrons, given by
\begin{eqnarray} 
\label{eq:KSpot}
v_{\rm eff}^{e\sigma}({\bf r})&=&\frac{1}{2}m^*\omega_0^2r^2+
v_{xc}^{\sigma}(n^e({\bf r}),\zeta^e({\bf r}))
+\int \frac{n^e({\bf r}\,')}{|{\bf r}-{\bf r}\,'|}\, d{\bf r}\,' \nonumber \\
&-&\int \frac{n^h({\bf r}\,')}{\sqrt{|{\bf r}-{\bf r}\,'|^2+z_0^2}}\, d{\bf r}\,'.
\end{eqnarray}
The electron density $n^e({\bf r})$ in Eq.~\ref{eq:KSpot} is given by
\begin{equation} \label{eq:n}
n^e({\bf r})=\sum_{i,\sigma }|\psi_{i\sigma}^e({\bf r})|^2,
\end{equation}
where $\sigma$ corresponds to the two spins 
$\sigma=(\downarrow,\uparrow)$ and $i$ refers to the lowest occupied states.
A corresponding relation holds for the hole density $n^h({\bf r})$
in the other dot.
$\zeta^e({\bf r})$ is the spin polarization for electrons, 
$\zeta ^e({\bf r})=(n^e_{\uparrow }({\bf r})-n^e_{\downarrow }({\bf r}))/n^e({\bf r})$.
The Kohn-Sham equations for electrons and holes were solved self-consistently using a
plane wave technique applied earlier to single 
dots~\cite{reimann1998,koskinen1997}. No symmetry restrictions for the
charge or  spin densities were applied.

For simplicity, let us assume equal effective masses of electrons and holes, 
$m_e^*=m_h^*=m^*$, 
and also restrict the numbers of electrons and holes confined in 
each of the wells to be equal, $N_e=N_h=N$.
(Throughout the paper, we use effective atomic units
with the energy measured in effective Hartree, 
${\rm Ha}^*=m^*e^4/\hbar^2(4\pi\epsilon_0\epsilon)^2$
and the distance in effective Bohr radii,
$a_B^*=\hbar^2(4\pi\epsilon_0\epsilon)/m^*e^2$.
With the effective mass $m^*$ and dielectric constant $\epsilon $,
the results then scale to the  value of the semiconductor material 
in question.) The  vertical double dot is then 
described by three parameters: the distance between the dots $z_0$,
the strength of the confining potential $\omega_0$
(taken to be equal for both dots), 
and the number of electrons and holes.
Due to the attractive interactions between the electrons and holes and since we 
assumed that $N_e=N_h=N$, 
the obtained ground state electron density necessarily 
equals that of the holes,
$n^e({\bf r})=n^h({\bf r})$; in the $z$-direction the densities 
are of course separated by $z_0$.
The spin density of the electrons and holes are also equal, but 
the relative direction of the electron and hole spins is not 
determined since there is no direct spin-spin interaction.

In the phase space determined by $z_0$ and $\omega_0$ 
three limiting cases are of particular interest:

(i) If both $z_0$ and $\omega_0$ are zero, the limit of the two-dimensional electron-hole
plasma is reached. 
The average density for large $N$ will approach the equilibrium density of the plasma.
In the case of small droplets, for non-closed shells the orbital degeneracies 
are removed by internal Jahn-Teller deformation of the density, 
resulting in intriguing geometries of the ground state densities. 
This regime was called earlier the ``ultimate jellium'' limit, and has 
also been studied 
in connection with the physics of simple metal clusters~\cite{deHeer1993} both 
in the two-dimensional \cite{reimann1998} and
in the three-dimensional case \cite{manninen1986,koskinen1995}.

(ii) When $\omega_0$ is still zero (or very small), but $z_0$ becomes larger, 
the average particle density in each
dot decreases and eventually, the particles begin to localize in a lattice.
In comparison to the formation of ``Wigner-molecules'' in quantum dots with 
electrons~\cite{dotwigner}, here
the localization is 
%strongly supported by 
clearly seen also in the local density approximation due to the 
the attractive interaction
between the electrons and the holes: they localize on top of each other forming
a lattice of bound electron-hole pairs (``localized excitons'').
Since the electrons and holes are confined in different 
layers, these electron-hole pairs have electric dipole moments and consequently 
repel each other. 
In each dot the localized electrons (or holes) show antiferromagnetic order 
(i.e., the so-called ``spin density wave''
~\cite{reimann1999,koskinen2001,manninen2001}) in the ground state.
The formation of bound electron-hole pairs is somewhat similar 
to the states discussed by Anisimovas and Peeters~\cite{anisimovas2002a}
for strong interdot coupling in a polarizing magnetic field. 

(iii) When $\omega_0$ increases the strong 
confinement eventually hinders the internal deformation
and the electron and hole densities are forced to be azimuthally symmetric.
The degeneracy in an open-shell dot can then be reduced by 
magnetization according to Hund's first rule \cite{tarucha1996,koskinen1997}.

Figure~\ref{fig:kimmof1} shows the phase diagram of the $N=4$ system. 
This case is particularly interesting since for a single
dot, the second shell is half-filled having two $p$-electrons and 
spin $S=1$~\cite{koskinen1997}.
In the limit of the ``ultimate jellium'' ($z_0=0$, $\omega_0=0$),
on the other hand,
the system is strongly deformed and non-magnetic.
Figure~\ref{fig:kimmof1} shows the regions of different magnetic and geometrical structures
in the $z_0$-$\omega_0$-plane. It is interesting to note that
for small interdot distances the deformed system is the ground state up to
a large value of $\omega_0$ (for $z_0=0$ the circular $S=1$ state becomes the ground state
when $\hbar\omega_0>~1.3$~Ha$^*$). For small values of $\omega_0$ we see a transition from the 
deformed state to the state with localized excitons when $z_0$ increases. 

A single dot with $N=6$ has a filled electronic shell and 
naturally $S=0$. Nevertheless, 
for a very weak confinement the ground state shows a spin-density wave~\cite{koskinen1997}.
A similar behavior is obtained for the electron-hole double dot at 
large values of the interdot distance $z_0$.
In the limit of small external confinement, the
ground state appears to be a hexagon of localized excitons. 
Classically, a more favorable geometry would be a pentagon with one exciton at
the center. However, in this case the hexagon is favored due to the 
formation of a spin-density wave, i.e., 
antiferromagnetic order 
of the spins in each dot (a centered pentagon would lead to ``frustration'').
The ground state of the $N=6$ double dot in the limit of the ``ultimate jellium''
($\omega _0=0, z_0=0$) does not have azimuthal symmetry, but is 
triangular, which seems to be the dominating 
geometry for closed-shell electron-hole clusters in two dimensions~\cite{reimann1998,reimann1997}.
The phase diagram (not printed here) shows a transition from a 
nonmagnetic triangular density to circular density when $\omega _0$ increases, and to localized electrons with a spin density wave when $z_0$ increases (for small $\omega _0$).
\begin{figure}
\includegraphics[width=0.9\columnwidth]{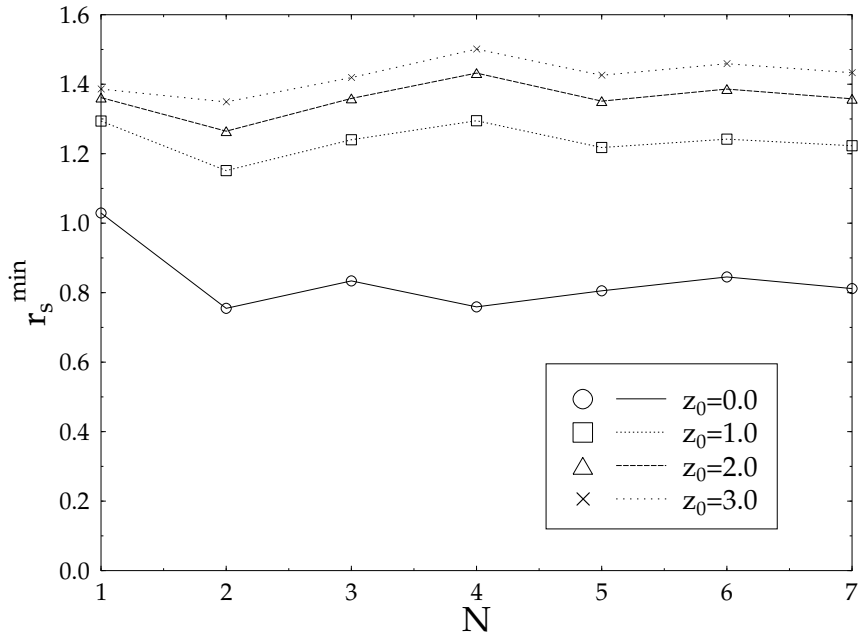}
\caption{Minimum of the {\it local} $r_s$ as a function of the number $N$ of 
electrons and holes for different values of the interdot distance $z_0$.
The average carrier density of a single dot was kept approximately 
independent of $N$ (see text).
} 
\label{fig:kimmof2}
\end{figure}
\begin{figure}
\includegraphics[width=0.9\columnwidth]{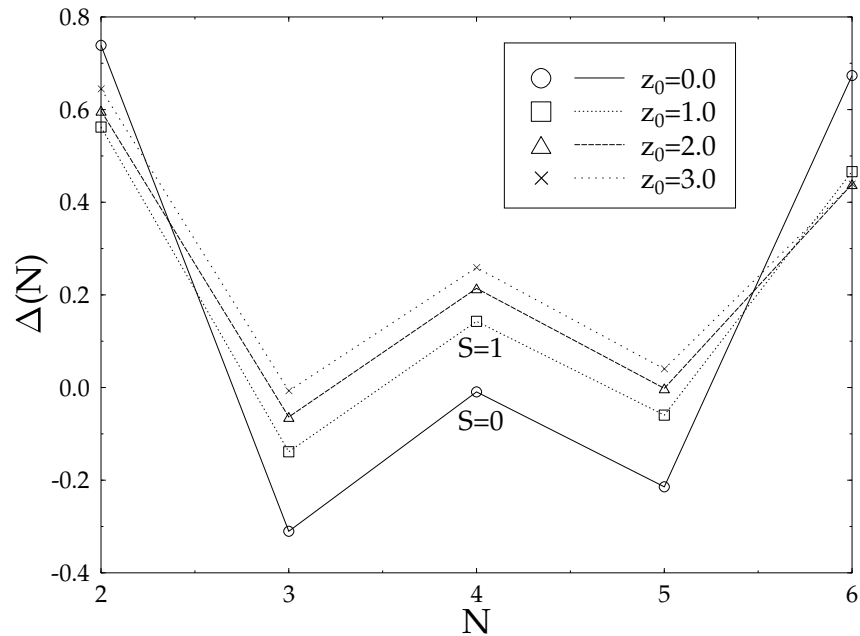}
\caption{The addition energy spectrum for different values of $z_0$.
The average carrier density of a single dot was kept approximately 
independent of $N$ (see text).}
\label{fig:kimmof3}
\end{figure}

In the above phase diagram, the number of electrons and holes $N$ was 
fixed. When $N$ is varied, however, experimental data on single quantum 
dots suggest~\cite{reimann2002} to keep the {\it average} density in each dot constant.
This can be achieved by the relation
$\omega_0^2(N)=e^2/(4\pi\epsilon\epsilon_0 m^*r_s^3\sqrt{N})$ 
~(see Ref.~\cite{koskinen1997}).
The parameter $r_s=1/\sqrt{\pi n}$ measures the in-layer 
coupling strength and is defined in terms of the in-layer areal density 
$n^{e(h)}({\bf r})$ for the electrons and holes, respectively. Here, we
use the $r_s$-value corresponding to the equilibrium
density of the two-dimensional electron gas ($r_s=1.51~a_B^*$).
Neglecting the electron-hole correlation, this 
density is also the  equilibrium density of the electron-hole plasma
without external confinement.

In the double dot system a single dot corresponds to the limit of infinite
$z_0$. When the interdot distance decreases, the electron-hole attraction increases the
effective confinement and the density increases slightly. However, since the $r_s$-value
has been chosen to be at the equilibrium density of the electron-hole plasma,
$r_s = 1.5~a_B^*$, 
 this effect is small.
Figure~\ref{fig:kimmof2} shows the minimum {\it local} value of $r_s$ (determined from the maximum density) 
obtained in the dots for 
different values of $N$ and $z_0$, 
using the above relation for $\omega_0(N)$. 
For $z_0=3~a_B^*$ the minimum $r_s$ is already
close to 1.5~$~a_B^*$ for all values of $N$. The variation as a function of $N$ is caused by the 
shell structure: The smallest value is obtained for the closed shell system, $N=2$,
and the maximum value for the half-filled shell $N=4$.
The lowest curve ($z_0=0$) corresponds to the situation where the deformation 
wins against Hund's rule, leading to a different dependence of the minimum $r_s$ on $N$.

In the case of single quantum dots 
the {\it addition energy spectrum} gives information about 
the shell structure. Maxima in the addition energy differences
were observed for closed shells, and at mid-shell spin alignment due to Hund's rule 
leads to enhanced stability~\cite{tarucha1996}. 
For the bipolar bilayer system studied here, 
we find that intriguingly, 
the addition energy spectra also reveal the internal
Jahn-Teller deformation of the system.
The difference in the 
electrochemical potentials of a bipolar double dot 
confining  $(N+1)$ and $N$ electrons 
{\it and holes} is given by $\Delta (N)=E(N+1)-2E(N)+E(N-1)$, i.e., 
the second differences of the corresponding total ground state energies 
$E(N)$.

Figure~\ref{fig:kimmof3}  shows the addition energy differences $\Delta (N)$
for $z_0\le 3~a_B^*$.
The addition energy maxima for the closed shells at 
$N=2$ and $N=6$ can be clearly seen.
An additional maximum occurs at mid-shell, $N=4$.
It is interesting to note that also the case $z_0=0$
shows a similar spectrum. In this case the maximum at $N=4$ is not 
caused by Hund's rule, but by the 
Jahn-Teller deformation, which is strongest at  half-filled shells.

In conclusion, we have studied symmetric vertical electron-hole double dots 
using the geometrically unrestricted spin-density functional method.
In the limit of small distances between the dots, the effective
confinement is governed by the electron-hole electrostatic attraction and 
the properties of the system are determined by
the Jahn-Teller deformation of the particle densities.
In the limit of weak external confinement, at increasing
interdot distance electrons and holes form separate ``excitons'' which localize
forming a lattice with antiferromagnetic spin-coupling in each dot. 
When the external confinement becomes large,
Hund's rule determines the spin in each dot.
The addition energy spectrum shows a maximum at half-shell.
At large interdot distances this maximum is caused by Hund's rule
while at small distances the maximum is a result of the Jahn-Teller
deformation of the particle density.

This work has been supported
by the Academy of Finland under the Finnish Centre of Excellence Programme
2000-2005 (Project No. 44875, Nuclear and Condensed Matter Programme at JYFL),
NORDITA, 
the Swedish Research Foundation (VR) and the Swedish Foundation for Strategic
Research (SSF).

% The Appendices part is started with the command \appendix;
% appendix sections are then done as normal sections
% \appendix

% \section{}
% \label{}

\end{document}